\pdfoutput=1
\documentclass{bmcart}

\usepackage{amssymb}
\usepackage{amsmath}%
\usepackage{amsfonts}%
\usepackage{graphicx}
\usepackage{pdflscape}
\usepackage{url}
\usepackage{ctable}
\usepackage{setspace}
\RequirePackage[compress]{natbib}
\usepackage[utf8]{inputenc} 

\startlocaldefs

\bgroup
\catcode`\.13

\egroup
\endlocaldefs

\begin{document}

\begin{frontmatter}

\begin{fmbox}
\dochead{Research Article}


\title{The surprising implications of familial association in disease risk}


\author[
   addressref={aff1,aff2},                   
   corref={aff1},                       
   email={morten.valberg@medisin.uio.no}   
]{\inits{M}\fnm{Morten} \snm{Valberg}}
\author[
   addressref={aff1},
   email={m.j.stensrud@medisin.uio.no}
]{\inits{MJ}\fnm{Mats Julius} \snm{Stensrud}}
\author[
   addressref={aff1},
   email={o.o.aalen@medisin.uio.no}
]{\inits{OO}\fnm{Odd O.} \snm{Aalen}}


\address[id=aff1]{
  \orgname{Oslo Centre for Biostatistics and Epidemiology, Department of Biostatistics, Institute of Basic Medical Sciences, University of Oslo}, 
  \street{POB. 1122, Blindern},                     %
  \postcode{N-0317}                                
  \city{Oslo},                              
  \cny{Norway}                                    
}
\address[id=aff2]{
  \orgname{Oslo Centre for Biostatistics and Epidemiology, Oslo University Hospital}, 
  \city{Oslo},                              
  \cny{Norway}                                    
}


\begin{artnotes}
\end{artnotes}

\end{fmbox}


\begin{abstractbox}

\begin{abstract} 
\parttitle{Background} A wide range of diseases show some degree of clustering in families; family history is therefore an important aspect for clinicians when making risk predictions. Familial aggregation is often quantified in terms of a familial relative risk ($FRR$), and although at first glance this measure may seem simple and intuitive as an average risk prediction, its implications are not straightforward.
\parttitle{Methods} We use two statistical models for the distribution of disease risk in a population: a dichotomous risk model that gives an intuitive understanding of the implication of a given $FRR$, and a continuous risk model that facilitates a more detailed computation of the inequalities in disease risk. Published estimates of $FRR$s are used to produce Lorenz curves and Gini indices that quantifies the inequalities in risk for a range of diseases.
\parttitle{Results} We demonstrate that even a moderate familial association in disease risk implies a very large difference in risk between individuals in the population. We give examples of diseases for which this is likely to be true, and we further demonstrate the relationship between the point estimates of $FRR$s and the distribution of risk in the population.
\parttitle{Conclusions} The variation in risk for several severe diseases may be larger than the variation in income in many countries. The implications of familial risk estimates should be recognized by epidemiologists and clinicians.
\end{abstract}




\end{abstractbox}
%

\end{frontmatter}



\section*{Background}
An important factor in the prediction of disease risk is the family history of disease, as the presence of such a history indicates that the patient has some underlying susceptibility. If few risk factors for a disease are known, but a familial association is observed, assessing family history of disease is one of the simplest and most cost-effective tools for risk prediction. However, the concept of familial risk is not as simple as it appears. Indeed, relatives may have an increased risk of disease due to genetic, epigenetic, common environmental/behavioral factors, or a combination of these. Very often these underlying determinants are unknown. Despite improved clinical health registries and the rapid development of genetic research methodology, observed factors explain only a minor proportion of the variation in disease risk within a given population \cite{aalen2014understanding}. For example, having a BRCA mutation increases the risk of breast cancer dramatically, but can explain only a minor proportion of all breast cancers \cite{balmain2003genetics}. Similarly, an underlying, unobserved heterogeneity in risk is likely to be important for many diseases \cite{aalen2014understanding}.

This failure to explain the causes of complex diseases is currently a matter of debate. For example, the impact of chance \textit{per se} in cancer development has been heavily discussed after being sparked by Tomasetti and Vogelstein, who implied that a large proportion of cancers are simply due to 'bad luck' \cite{tomasetti2015variation,Stensrud201783}. When the causes of a disease are unknown, studying its familial aggregation may help us understand how the risk is distributed in the population due to both observed and unobserved factors. This is not only important from a scientific point of view; it can also reveal the consequences of having an affected relative. In practice, this information may be important for genetic counseling and for follow-up of individuals with a family history of disease \cite{riley2012essential}. 

Familial associations are often quantified in terms of familial relative risks ($FRR$s). Generally, the $FRR$ denotes the risk of disease when a family member is affected compared to the risk level in the general population. Specific types of familial relationships, like first-degree relatives, parent-child, or siblings might also be of interest.  A familial association has been demonstrated in a wide range of diseases. In the last decades, $FRR$s for virtually all cancers have become readily available in the literature \cite{frank2015population}. For breast, colon, and prostate cancer, the risk has been reported to double when a family member (first-degree relative) has the disease \cite{frank2015population,jasperson2010hereditary, ripperger2009breast,johns2003systematic}, and the risk further increases if several family members are affected. Furthermore, several autoimmune and neurodegenerative diseases also have a substantial familial association \cite{hemminki2009familialALS,marder1996risk}. Therefore, understanding the information that is carried by an $FRR$ is becoming increasingly important.

Fundamentally, we can assume two different points of view when studying familial associations. One view consists of focusing on observed familial associations in incidence, e.g., measured as the FRR, which can be calculated immediately from the data. The other focuses on how the risk varies between families, i.e., rather than consider summary measures like the FRR in isolation, we can investigate how the disease risk is distributed across families in the population. Indeed, it is logical that a familial association of disease risk implies a variation between individuals in a population. However, the connection between these two views is not immediate or intuitively easy. It is generally under-appreciated that a risk factor that is correlated within a family has to be very strong to produce even a moderate familial association in disease risk \cite{khoury1988can,aalen1991modelling,hopper1992familial}. In the present study, we will use different models to illustrate various possible, potentially surprising, relationships between these two views. 

First, we will study a simple dichotomous risk model, by dividing the population into two distinct risk groups. All members of a family (e.g., a group of siblings) belongs to the same risk group. This model provides simple, yet informative illustrations of the relationship between observed familial risk and actual differences in disease risk. If $FRR$s are available for both one and two affected family members, then the actual relative risk between the two risk strata, as well as the size of both of these strata, can be calculated.  Next, we will study a slightly more detailed model, using a continuous distribution for the risk of developing disease. This facilitates the computation of Lorenz curves and Gini coefficients, which are well established methods for measuring inequality in wealth in economics. Finally, we discuss the issues highlighted in the paper, and their implications.

\section*{Methods}

\subsection*{Dichotomous risk model}
In the dichotomous risk model, we will consider a population that is divided into two groups: a high-risk group and a low-risk group. The risk is assumed to be the same for all individuals in the same group, and all individuals from the same family belongs to the same risk group. Thus, the type of familial relationship is not considered here. This is a simplification, but provides convenient and illustrative examples.

The probabilities of belonging to the high- and low-risk groups are $q$ and $1-q$, respectively. The risk of acquiring a disease, say within a given age, is $p_{h}$ and $p_{l}$ within these two groups. The individual relative risk (IRR), comparing a high-risk individual with a low-risk individual is $IRR = p_{h}/p_{l}$.

The $FRR$ is calculated by considering two people from the same family. That is, assuming one family member acquired the disease at a given age, we may calculate the probability that the other family member will acquire the same disease by the same age. This probability is then divided by the average risk of disease:

{\scriptsize
\begin{equation}
\begin{split}
FRR_1 &=\frac{Prob(\mathrm{person\ develops\ disease}|\mathrm{family\ member\ has\ disease})}{Prob(\mathrm{person\ develops\ disease})}\\ &=\frac{qIRR^2+1-q}{(qIRR+1-q)^2}.%
\end{split}
\label{FRR1}
\end{equation}
}
This formula can be expanded to include a larger family, whose members might have different disease statuses. For a family with three members, assuming two have acquired the disease, the $FRR$ is calculated as
{\scriptsize
\begin{equation}
\begin{split}
FRR_2 &=\frac{Prob(\mathrm{person\ develops\ disease}|\mathrm{two\ family\ members\ has\ disease})}{Prob(\mathrm{person\ develops\ disease})}\\ &=\frac{qIRR^3+1-q}{(qIRR+1-q)(qIRR^2+1-q)}.%
\end{split}
\label{FRR2}
\end{equation}
}
Both $FRR_1$ and $FRR_2$ depend on the $IRR$ and the size of the risk groups ($q$). However, given the $IRR$, the $FRR$s are independent of the actual disease risks in each group, $p_{h}$ and $p_{l}$.

\subsection*{Continuous risk model}
A more detailed description of the risk distribution can be obtained by considering a risk that varies continuously in the population. Let $P$ be the probability of acquiring a disease for a given person over a given time interval (say that the disease occurs before a certain age). This probability is a random variable in that 1) we let $P$ vary between families, but assume it is identical for members of the same family; 2) $P$ accounts for all variability in risk between families, hence individuals from the same family are independent given the family's risk level. Define $E[P]$ as the expected value of $P$, i.e., the mean risk of acquiring the disease. Thus the risk of acquiring the disease if one family member is affected is calculated as
\begin{equation*}
\begin{split}
&Prob(\mathrm{person\ develops\ disease}|\mathrm{family\ member\ has\ disease})\\
&=\frac{Prob\mathrm{(both\ family\ members\ develop\ disease})}{Prob\mathrm{(family\ member\ develops\ disease})} \\
&=\frac{E[Prob\mathrm{(both\ family\ members\ develop\ disease|familial\ risk\ level\ } P)]}{E[Prob(\mathrm{person\ develops\ disease|familial\ risk\ level\ } P)]} = \frac{E[P^2]}{E[P]}.
\end{split}
\end{equation*}
Here, $E[P^{2}]$ is the mean risk of two individuals from the same family acquiring the disease over a given time interval. By dividing by the mean risk of acquiring the disease, $E[P]$, the $FRR$ is found as
\begin{equation}
FRR=\frac{E[P^{2}]}{E[P]^{2}} =1+\frac{Var[P]}{E[P]^{2}}=1+CV^{2},
\label{FRR-CV}
\end{equation}
where $CV$ is the coefficient of variation of $P$ \cite{moger2004analysis,aalen2014understanding,ABG}. Since $P$s distribution is nonnegative, even a moderate value of $CV$ implies a skewed distribution.

In the examples to come, we will assume a beta distribution for $P$ \cite{smith1971recurrence}, which is a very flexible family of distributions. The density on $[0,1]$ is given by%
\[
f(p;\alpha,\beta)=\frac{p^{\alpha-1}(1-p)^{\beta-1}}{B(\alpha,\beta)}
,\qquad\alpha>0,\beta>0
\]
where $B(\alpha,\beta)$ is the beta function. The expectation of the distribution is given by
\begin{equation}
\mathrm{E}[P](\alpha,\beta)=\frac{\alpha}{\alpha+\beta}.
\label{expected}
\end{equation}
The squared coefficient of variation is given as
\begin{equation}
\mathrm{CV}^{2}[P](\alpha,\beta)=\frac{\beta}{\alpha(\alpha+\beta+1)}.
\label{CV2}
\end{equation}

We wish to measure and understand the skewness of the distribution, that is, how (un)evenly the risk is distributed across the population. For this, we use published estimates of $FRR$s and the life-time risks of the diseases in question, and plug them into Expressions \eqref{FRR-CV}, \eqref{expected} and \eqref{CV2}.

\subsubsection*{The Lorenz curve highlights the inequality in risk}
In parallel to epidemiological studies of the distribution of disease risk, economic studies have a strong interest in the distribution of wealth. In economic studies, the Lorenz curve and the Gini index are important measures. The Lorenz curve often represents the percentage of income in the population earned by a certain cumulative percentage of the population. If the Lorenz curve is a straight diagonal line, it represents perfect equality in income. The Gini index is calculated as the area between the theoretical perfect equality line and the actual Lorenz curve (which is usually less than $0.5$), multiplied by two. Thus, if the Gini index is 0, there is a perfect equality in income. Despite its usefulness, the Lorenz curve has been much less used in epidemiological studies \cite{mauguen2016using}. In studies of disease, the Lorenz curve displays the expected proportion of disease burden carried by the X\% at lowest risk. Intuitively, in a medical setting the Gini index denotes the deviation from equality in risk, i.e., the Gini index is equal to 0 if all families have the same disease risk, and the index is 1 if only one family is at risk. That is, the larger the Gini index, the larger the deviation from equality.

\section*{Results}
\subsection*{Dichotomous risk model: Moderate $FRR$s can imply large differences in risk}
\label{dichotomous}
\subsubsection*{\\Small high-risk groups produce large $FRR$s}
Figure \ref{FRRq=0.01}a shows both $FRR_1$ and $FRR_2$ plotted against the $IRR$ in a situation where only $q=1\%$ of the population belongs to the high-risk group, and the $IRR$ varies from $1$ to $20$. In this situation, the $FRR$s are much lower than all the $IRR$ values. Assuming one affected family member, a doubling of the $FRR$ requires an $IRR$ of $12.2$. Assuming two affected family members leads to a better correspondence between the $FRR$ and $IRR$, but the $IRR$ still has to be larger than 5 to produce a doubling of the $FRR$. It is perhaps more natural to consider the $IRR$ as a function of the $FRR$, since the latter is observable in practice. Such a plot is given in Figure \ref{FRRq=0.01}b, in which the discrepancies between the two relative risks when one or two family members are affected are even more pronounced.

Figure \ref{FRRq=0.01} revealed two important characteristics. First, the $FRR$ can be substantially different from the $IRR$. Closely related to this example, Peto et al. described a mendelian context with a 10-fold increase in the disease risk among carriers of a dominant genotype (i.e. an $IRR = 10$), but having a sibling with the disease only doubled the risk (i.e. a $FRR = 2$) \cite{peto1980banbury}. Second, the $FRR$ depends strongly on the number of affected family members, in particular for rare diseases. One example is testicular cancer, for which the $FRR$ is 6 given one affected brother, and increases to almost 22 given two affected brothers \cite{valberg2014hierarchical}. Intuitively, if the disease is rare, having an increasing number of affected family members will increase the likelihood of the family actually belonging to the (small) high-risk group of the population. Notably, the $FRR$ would be reduced if there were additional disease free members in the family. For rare diseases, however, the main determinant of the $FRR$ is the number of affected family members, as has been shown for testicular cancer \cite{valberg2014hierarchical}.

\subsubsection*{$FRR$s may be misleading for common diseases.}
Figures \ref{FRRq=0.5and0.8}a and \ref{FRRq=0.5and0.8}b display the relationship between the $FRR$ and the $IRR$ when $q=0.5$. That is, when we assume that 50\% of the population belong to the high-risk group. $FRR$s assuming one or two affected family members are below 2 irrespective of the size of the $IRR$. This is caused by the fact that the disease appears quite frequently in the population, therefore having a family member with disease does not provide much information on whether the family is at high or low risk. Although having more affected family members will increase this likelihood, Figures \ref{FRRq=0.5and0.8}a and \ref{FRRq=0.5and0.8}b show that this increase is not necessarily substantial. Figures \ref{FRRq=0.5and0.8}c and \ref{FRRq=0.5and0.8}d show a situation in which the high-risk group accounts for 80\% of the population ($q=0.8$), i.e., a minority of 20\% has lower disease risk. The $FRR$ does not exceed 1.25 in this situation, irrespective of the size of the $IRR$. 

Figure \ref{FRRirr=20} presents the relationships between $FRR$s, assuming one or two affected family members, and the proportion of high-risk individuals $q$, given an $IRR=20$. The curves peak at very low values of $q$, illustrating a better correspondence between the $FRR$ and the $IRR$ when the high-risk group is small. The plots of the $IRR$ in terms of the $FRR$ (Figures \ref{FRRq=0.01}b, \ref{FRRq=0.5and0.8}b, and \ref{FRRq=0.5and0.8}d) clearly illustrated that considering the $FRR$ alone can give a very misleading impression of the true risk distribution in the population, especially for common diseases. 

\subsubsection*{Inferences based on published $FRR$s}
For several diseases, estimates of both $FRR_1$ and $FRR_2$ are available in the literature. Expressions \ref{FRR1} and \ref{FRR2} can then be solved to find both the corresponding $IRR$ and high-risk proportion, $q$. In Table \ref{solveBoth}, these values are given for seven selected cancers. For testicular cancer, the $FRR_1=5.9$ and $FRR_2=21.7$ translate into a high-risk group consisting of $1\%$ of the population, which has 30.6 times the risk of the remaining $99\%$ of the population. For breast cancer, $FRR_1=1.8$ and $FRR_2=2.9$ shows that the $10\%$ of the population that make up the high-risk group has on average 5.2 times the risk of developing the cancer compared with the remaining 90\%. In the dichotomous model, the IRR represents the ratio of the average risk in the two groups. This is a crude simplification, but Table \ref{solveBoth} illustrates that it is possible to gain useful information on how risk is distributed in a population, even when assuming a dichotomous model. However, rather than being dichotomous, the real risk distribution is likely continuous and skewed \cite{aalen2014understanding}.

\subsection*{Continuous risk model: Large inequalities in individual risk are likely}
\label{sec:2}
The results from the dichotomous model gave an intuitive understanding of the challenges of handling familial risks. However, in real life, the individual risk may often vary continuously across the population. This reflects the fact that the individual risk of most diseases is caused by a combination of several inherited and environmental factors.

\subsubsection*{The risk of severe diseases varies more than income in the USA}
Parkinson's disease has a complex etiology, and many factors, both heritable and non-heritable, could contribute to disease risk. We therefore consider the risk to vary continuously across the population. The life-time risk of Parkinson's disease is approximately 1\% (i.e. $E[P]=0.01$) and the $FRR$ has been estimated at 2.3 \cite{marder1996risk}. We assume that the beta distribution can describe the risk distribution, and use the approach described above to make inferences about its shape. Solving Equations \eqref{FRR-CV}, \eqref{expected} and \eqref{CV2}, $\alpha=0.75$ and $\beta=74$ are obtained. The corresponding Lorenz curve is shown in Figure \ref{lorentz}a, and the Gini index equals 0.55, which implies a very large variation in individual risk. This is further demonstrated in Figure \ref{lorentz}b, which shows a Manhattan-type skyscraper landscape, with a large variation in the height of the columns. These heights are simulations from the distribution and resemble the disease risk of different families. 

For Parkinson's disease, the large variation in risk is not obvious from its moderate $FRR$. Indeed, many severe cancers could share similar patterns. For example, in the USA, cancer of the pancreas (life-time risk of 1.5\% and a $FRR$ of 2.19), leukemias (life-time risk of 0.96\% and a $FRR$ of 2.01), and cancer of the stomach (life-time risk of 1.78\% and a $FRR$ of 1.92) all show these features \cite{howlader2011seer,frank2014population}, producing Gini indexes of 0.54, 0.50 and 0.49, respectively. 

A further increased $FRR$ would imply an even more skewed risk distribution for the same level of life-time risk. In Figure \ref{lorenz3eks}, we display the risk distribution and the Lorenz curves when the lifetime risk is $E[P]=1\%$, for different $FRR$s. In this scenario, even a moderate $FRR$ of 1.5 implies a distribution that is considerably skewed; the Lorenz curve corresponds to a Gini index of 0.37, and the 10\% with the highest risk accounts for 26\% of the diagnoses. Increasing the $FRR$ to 6 yields a Gini index of 0.80, and the 10\% with the highest risk accounts for 65\% of the diseased. The latter example is similar to testicular cancer, which has a reported $FRR$ of 5.88 and a life-time risk of $0.9\%$ \cite{valberg2014hierarchical,CiN20015}.

Breast cancer has a reported life-time risk of 12\% and a FRR of 1.8 \cite{desantis2016breast,collaborative2001familial}, which renders a Gini index of 0.47. The 10\% with highest risk accounts for 30\% of the diagnoses, and the mean risk in that 10\% is 6.2 times the mean risk in the remaining 90\%. This is similar to what was found in Table \ref{solveBoth} and to hypothesized risk distributions for breast cancer \cite{balmain2003genetics,chatterjee2016developing}. 

More common diseases may also have a considerable variation in risk. For example, diabetes type 2 has a life-time risk of more than $30\%$ in the USA \cite{narayan2003lifetime} and is estimated to have a $FRR$ of 2.24 \cite{weires2007familiality}. This yields a Gini index of 0.60, again implying a strong heterogeneity in the risk of the disease that is not reflected by the moderate $FRR$. For this example, it is also intuitive: We know that several risk factors, e.g., body weight, physical inactivity and particularly genetic factors, are unequally distributed between families. Evaluating such factors clearly gives a better indication of the disease risk than the $FRR$.

As an interesting comparison, the Gini index for income distributions in the Scandinavian countries ranges from 0.25 to 0.27, in the USA it is 0.45, while Lesotho tops the list at 0.63 \cite{cia}. In other words, all the diseases mentioned above show a variation in individual risk that is larger than the variation in income in the USA. 

\subsubsection*{Extreme risk in the highest percentile}
The results from the dichotomous risk model and the graph in Figure \ref{FRRirr=20} suggested that large $FRR$s tend to be found for rare diseases. Diabetes type 1 has a prevalence of about 0.2\% among Americans under 20 years of age and an $FRR=12$ among singleton siblings \cite{hemminki2009familial}. Diabetes type 1 occurs when the immune system targets the insulin producing cells in the pancreas. This processes occurs in genetically susceptible individuals, but it is postulated that environmental effects, such as viral infections, could contribute to developing the disease. If we consider the risk of developing diabetes to be $E[P] = 0.2\%$, we obtain a Gini index of 0.89. In our model, 25\% of the individuals with disease would belong to a group of high-risk families comprising only 1\% of the total population. The median risk among this 1\% of high-risk families would be more than 10,000 times higher than the 99\% remaining population.

\section*{Discussion}
\label{sec:4}
Even when a familial disease association is modest, the variation in individual risk could be substantial. Using simple statistical models we have shown relationships between the $FRR$ and the $IRR$ that are intuitively surprising. Even familial disease risks that are apparently modest, like the doubling of risk in relatives of patients with breast cancer or colon cancer, imply large variations in risk between families \cite{byrnes2008so,hopper2011disease}. It is important that clinicians and epidemiologists are able to recognize the consequences of these relationships. In particular, even if an $FRR$ is modest, some families would have a remarkably larger risk of developing the disease than others. In fact, the risk distribution may be more skewed than the income distributions of many countries. This points to a fundamental, and to a large extent unexplained, variation in risk. This variation occurs due to genetic predispositions, but also environmental, and various other sources of variation, including more unspecific, random variation \cite{aalen2014understanding}. Importantly, this unexplained variation in risk may lead to considerable selection bias in observational studies \cite{stensrud2017a,stensrud2017b}. Estimates of FRRs may actually be used to adjust for this bias, e.g., in Cox proportional hazards models \cite{stensrud2017handling}.  

Using $FRR$s for counseling is tempting when other tools for predicting individual risks, such as genetic tests or biomarkers, are lacking. It could also be tempting to use $FRR$s to aid in targeted screening \cite{hemminki2014collection,frank2015population}. For example, many countries use family history of colorectal cancer in screening recommendations today \cite{win2014risk}. However, the quantifications of familial associations are often crude, sometimes limited to one number for a given familial relationship. We emphasize that this $FRR$ could be deceiving, even for relatively rare diseases. A correct interpretation of the $FRR$ is therefore crucial. Individuals could suffer from unnecessary worry and testing if the $FRR$ is misinterpreted. On the other hand, we could also fail to identify high-risk individuals by putting too much confidence on moderate $FRR$s, which are often population averages. In reality, $FRR$s vary continuously along a number of parameters. The number of affected family members is of obvious importance, but the number of healthy members may also provide important information. Furthermore, the aspect of \textit{time} is essential. The ages at which family members acquire a disease (or remains disease free) is crucial to determining the level of risk. Taking these aspects into account is possible, e.g., by using methods from survival analysis that provides the opportunity to make risk predictions based on the family history of specific individuals \cite{moger2011hierarchical,valberg2014hierarchical,ABG}. These detailed $FRR$s could be more useful for individualized risk predictions, and could thus contribute to targeted screening as well as help focus preventive efforts.

Estimated $FRR$s from epidemiological studies are often taken at face value in other biomedical disciplines. For example, these estimates are frequently used as reference values in genome-wide association studies (GWAS') \cite{houlston2008meta,houlston1996genetics,cox2007common,witte2014contribution}, in which the proportion of the $FRR$s that can be attributed to specific, or all known risk alleles are frequently reported. However, methods for estimating $FRR$s may vary in their complexity. Thus, the proportion of the $FRR$ explained in GWAS' is dependent on the methods used in the epidemiological studies they base their reference on. A review of the genetics of type 1 diabetes stated that 34 susceptibility loci explained 60\% of the variation in risk, based on an estimated $FRR$ reference value of 15 \cite{polychronakos2011understanding}. The review was criticized for using this reference value as it was estimated in a period when risk was lower than it is presently, and it was suggested that 12 was a better estimate \cite{hemminki2012familial}. Applying this reference value changed the explained variance to 75\%. The criticism was sound, but alternative methods for estimating the $FRR$ might have produced a different (or more detailed) estimates \cite{moger2011hierarchical,valberg2014hierarchical,ABG}. Rather than focusing on a single estimate of the $FRR$, investigating the sensitivity to different methods and measures of association seems like a good practice. Furthermore, the $FRR$ encompasses much more than merely genetic inheritance. If all susceptibility alleles could be identified, they would not explain 100\% of the familial risk. Also, a gene with a strong impact on disease susceptibility might only explain a minor fraction of the $FRR$, depending on the magnitude of the $FRR$ and the population frequency of the allele \cite{rybicki2000relationship}. Hence, it is not clear what information that is provided by this measure (proportion of $FRR$ explained), and it seems difficult to compare it across different diseases.

Interpreting familial risks can be challenging, especially when relating them to heredity. Even if a mutation gives a high life-time risk of, say 50\%, half of the high-risk individuals would not be expected to develop the disease. Furthermore, the mutation would only be passed on to half of the individuals in the next generation, on average. Consequently, many cancers diagnosed in genetically predisposed individuals would appear to be sporadic (i.e., occurring in individuals without a family history). Such an argument has been used to suggest that the majority of hereditary, early-onset breast cancers appears sporadically in the population \cite{cui2000majority}. In a recent study, Cremers et al. found that the same single-nucleotide polymorphisms (SNPs) were associated with an increased risk of both hereditary (defined as three cancers in first-degree relatives) and sporadic prostate cancer, and concluded that "hereditary prostate cancer most probably is merely an accumulation of sporadic prostate cancers" \cite{cremers2015known}. Although that might be the case, it is possible to draw the opposite conclusion; that sporadic cancers are, in fact, also inherited. This underscores the fact that relating the $FRR$ to heredity is not straightforward. If assessing the degree of heritability is the aim, one should look to other types of studies, e.g., to twin studies \cite{mucci2016familial,stensrud2017inequality}. However, the large variation in risk implied by even a moderate familial risk is unlikely to be explained by environmental risk factors correlated in families \cite{khoury1988can,aalen1991modelling,risch2001genetic}.

\section*{Conclusions}
We have seen that the interpretation of the $FRR$ depends on the context. First, the life-time prevalence of the disease is important. The specific definition of the $FRR$, in particular the familial relationships that are studied, may also be crucial. Even if a precise definition of the $FRR$ is given, it may be estimated in several different ways. If the $FRR$ is calculated to perform risk predictions and genetic counseling, a detailed $FRR$ based on individual characteristics (i.e. familial histories) is desirable. In any case, we have shown that even simple $FRR$s, averaged over the population, can reveal important information on how the risk is distributed in the population. Even a moderate $FRR$ may imply a very skewed risk distribution.



\begin{backmatter}

\section*{Abbreviations}
CV: Coefficient of variation; FRR: Familial relative risk; GWAS: Genome-wide association study; IRR: Individual relative risk; SNP: single-nucleotide polymorphism.

\section*{Competing interests}
The authors declare that they have no competing interests.

\section*{Author's contributions}
All authors (MV, MJS and OOA) contributed to the conception of the study, the drafting of the manuscript, revising the manuscript for important intellectual content and approved the final version of the manuscript.

\section*{Acknowledgements}
We would like to thank Steinar Tretli and Tom Grotmol at the Cancer Registry of Norway for valuable discussions and comments on an earlier version of the manuscript.

\section*{Funding}
This work was partially supported by the Norwegian Cancer Society, grant number 4493570, and the Nordic Cancer Union, grant number 186031.

\section*{Ethics approval and consent to participate}
Not applicable.

\section*{Consent for publication}
Not applicable.

\section*{Availability of data and material}
Calculations were performed based on estimates available in the references given in the text.


\bibliographystyle{vancouver} 
\bibliography{famArtBMC}      




\section*{}

\begin{landscape}
\begin{figure}[]
 	\centering
  	\includegraphics[scale=0.7]{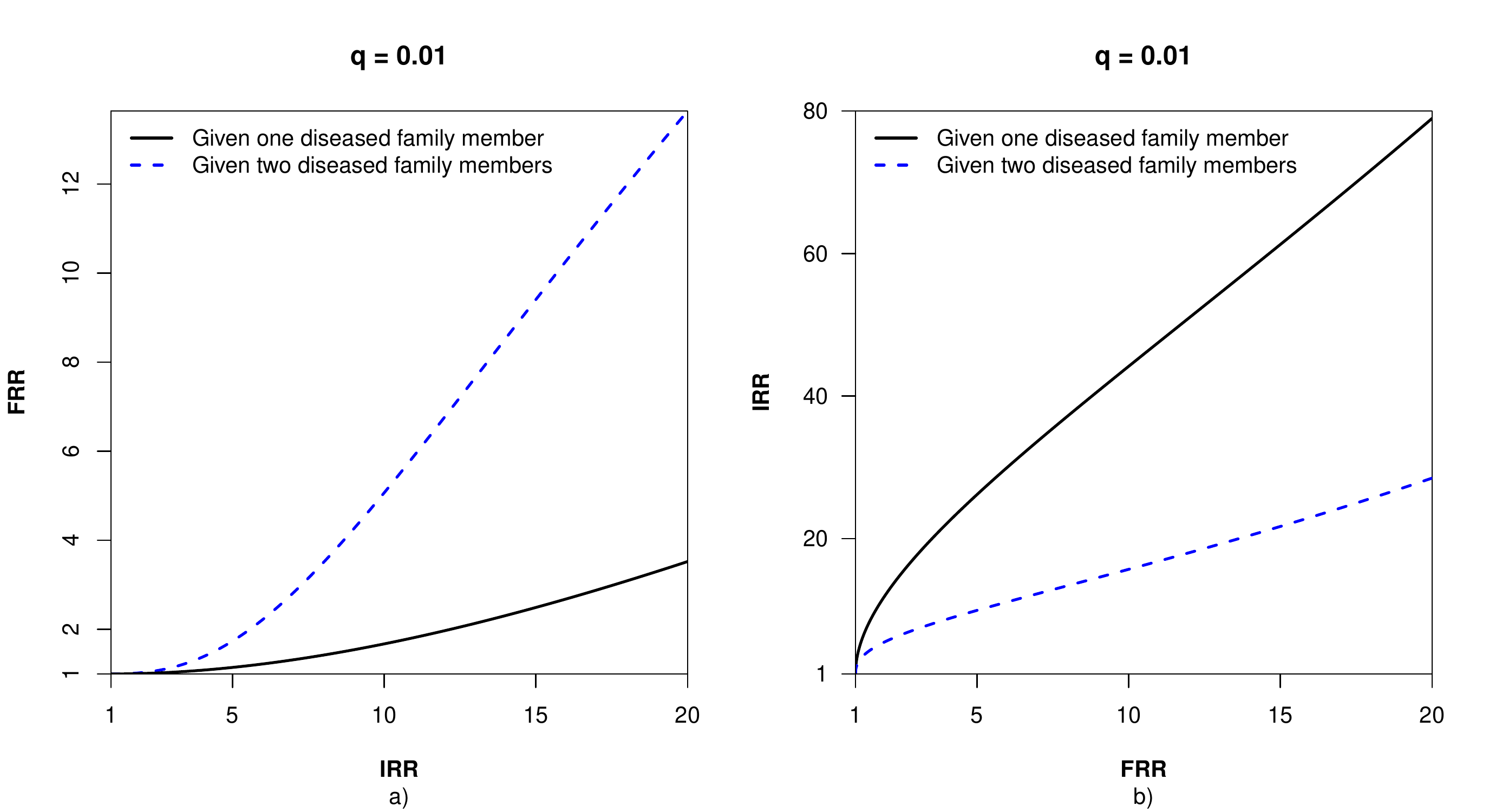}
 	\caption{a) The familial relative risk ($FRR$) as a function of the individual relative risk ($IRR$) defined in terms of the dichotomous model in Expression \eqref{FRR1} and \eqref{FRR2}, assuming one or two diseased family members, respectively. b) The $IRR$ as a function of the $FRR$. 1\% of the population belong to the high-risk group ($q=0.1$) in both panels.}
 	 \label{FRRq=0.01}
\end{figure}
\end{landscape}

\begin{landscape}
\begin{figure}[]
 	\centering
	\includegraphics[scale=0.7]{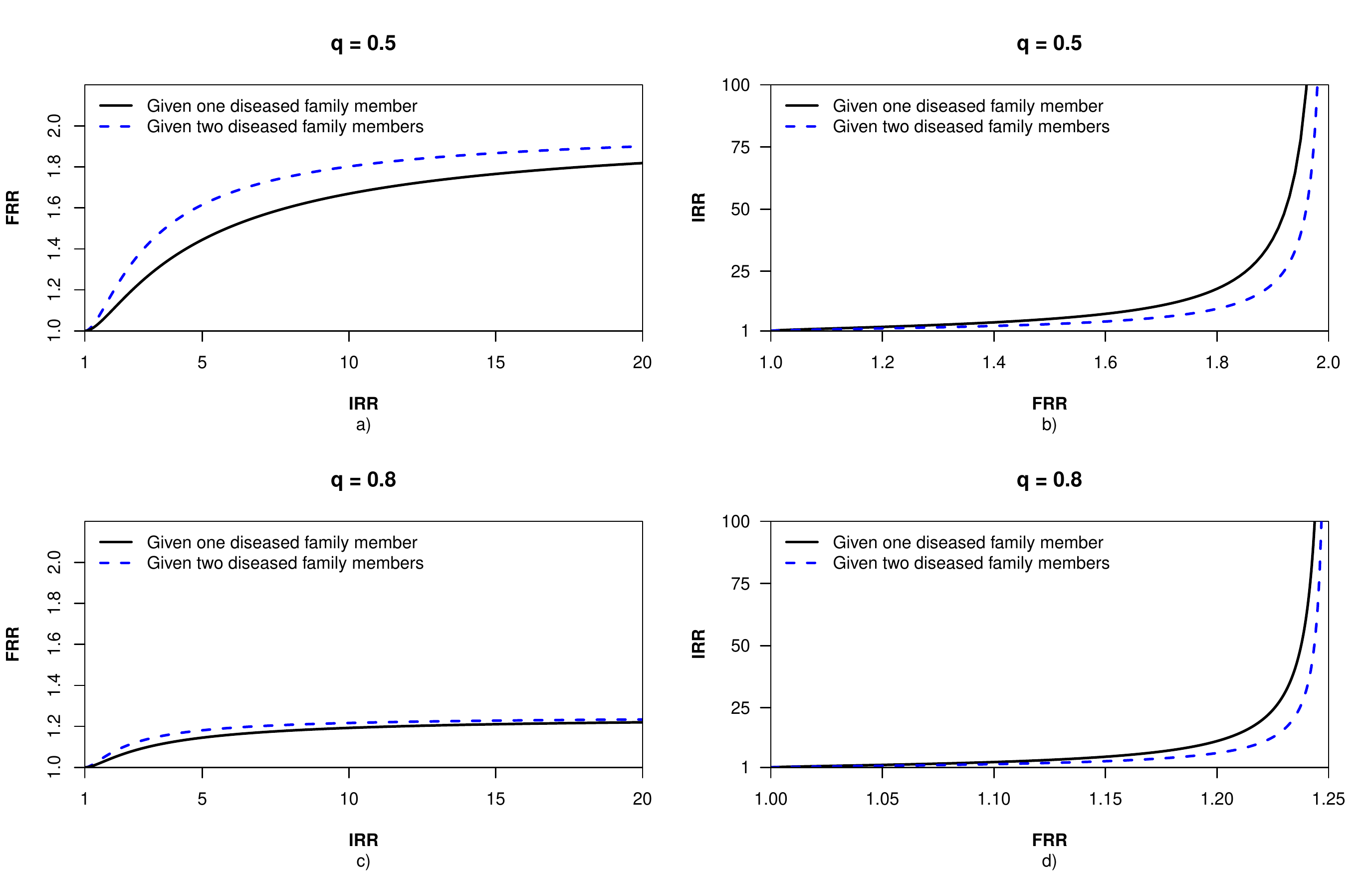}
 	\caption{a, c) The familial relative risk ($FRR$) as a function of the individual relative risk ($IRR$) defined in terms of the dichotomous model in Expression \eqref{FRR1} and \eqref{FRR2}, assuming one or two diseased family members, respectively. b, d) The $IRR$ as a function of the $FRR$. 50\% ($q=0.5$) and 80\% ($q=0.8$) of the population belong to the high-risk group in panels a, b) and c, d), respectively.}
 	 \label{FRRq=0.5and0.8}
\end{figure}
\end{landscape}

\begin{figure}[]
 	\centering
  	\includegraphics[scale=0.8]{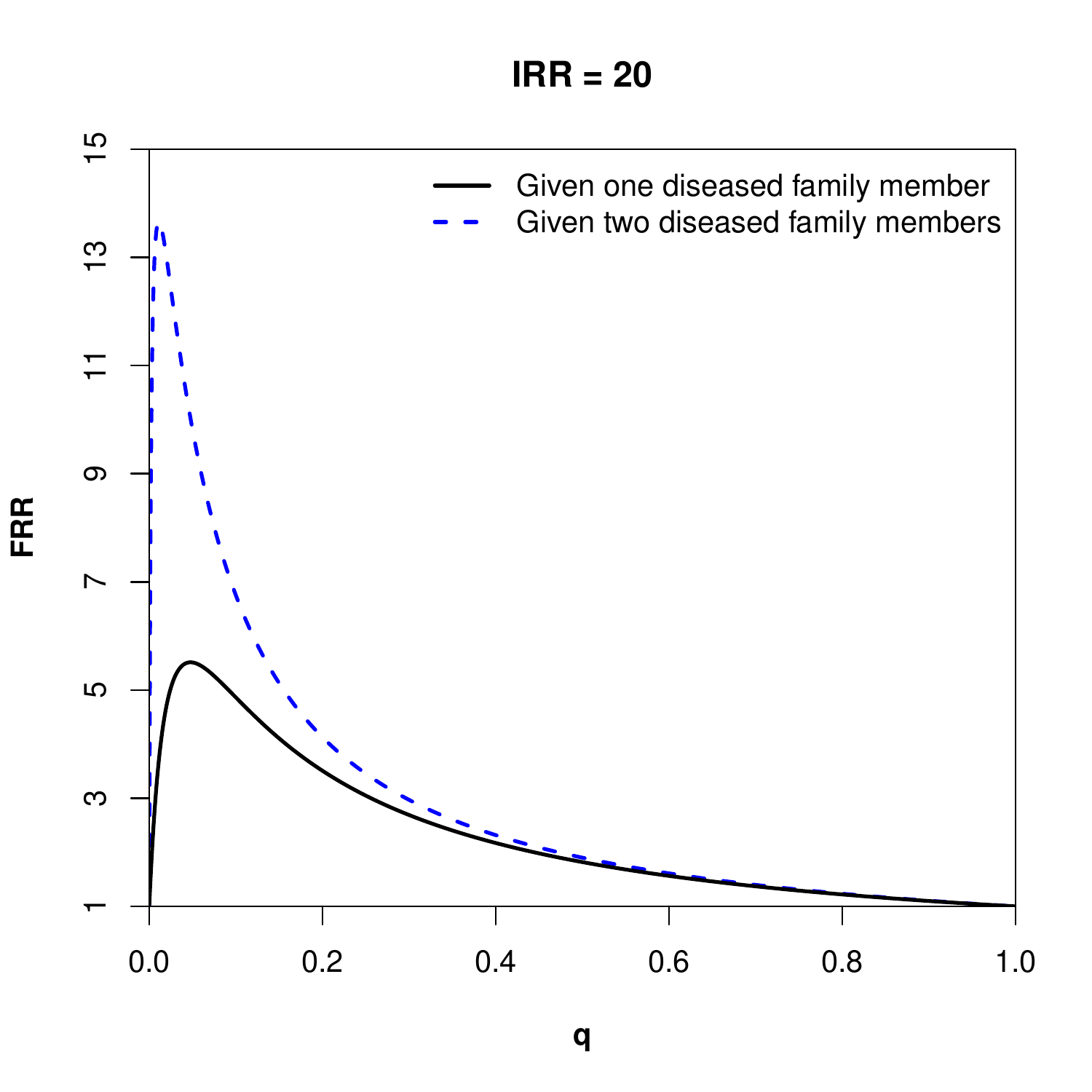}
 	\caption{The familial relative risk ($FRR$) as a function of the proportion, $q$, of the population belonging to the high-risk group defined in terms of the dichotomous model in Expression \eqref{FRR1} and \eqref{FRR2}, assuming one or two diseased family members, respectively. The individual relative risk ($IRR$) is 20.}
 	 \label{FRRirr=20}
\end{figure}

\begin{landscape}
\begin{figure}[]
 	\centering
	\includegraphics[scale=0.7]{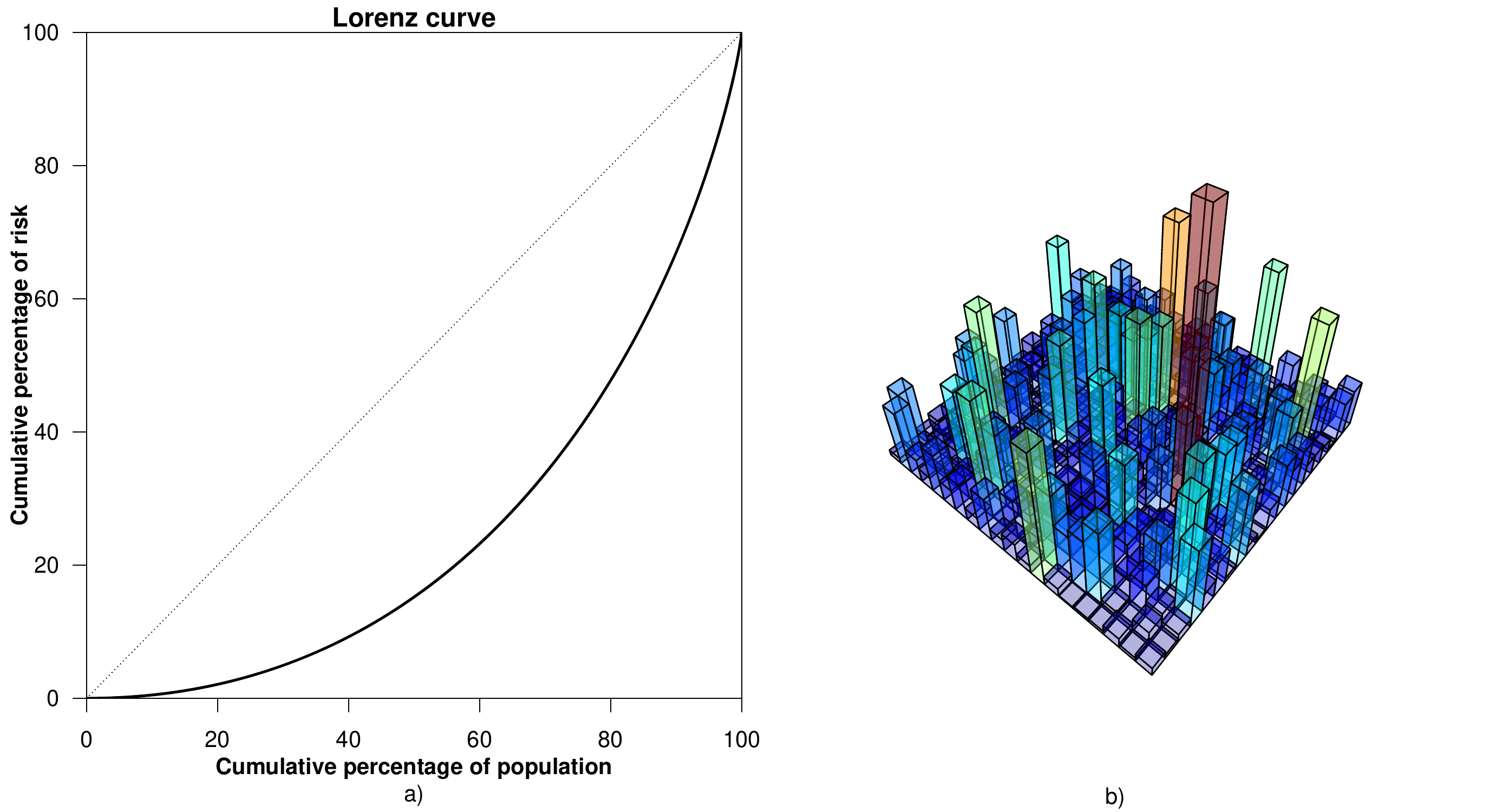}
 	\caption{The Lorenz curve (a) and 400 samples (b) from a beta distribution resulting from a familial relative risk of 2.3 and $E[P]=0.01$, representing Parkinson's disease.}
 	 \label{lorentz}
\end{figure}
\end{landscape}

\begin{landscape}
\begin{figure}[]
 	\centering
	\includegraphics[scale=0.7]{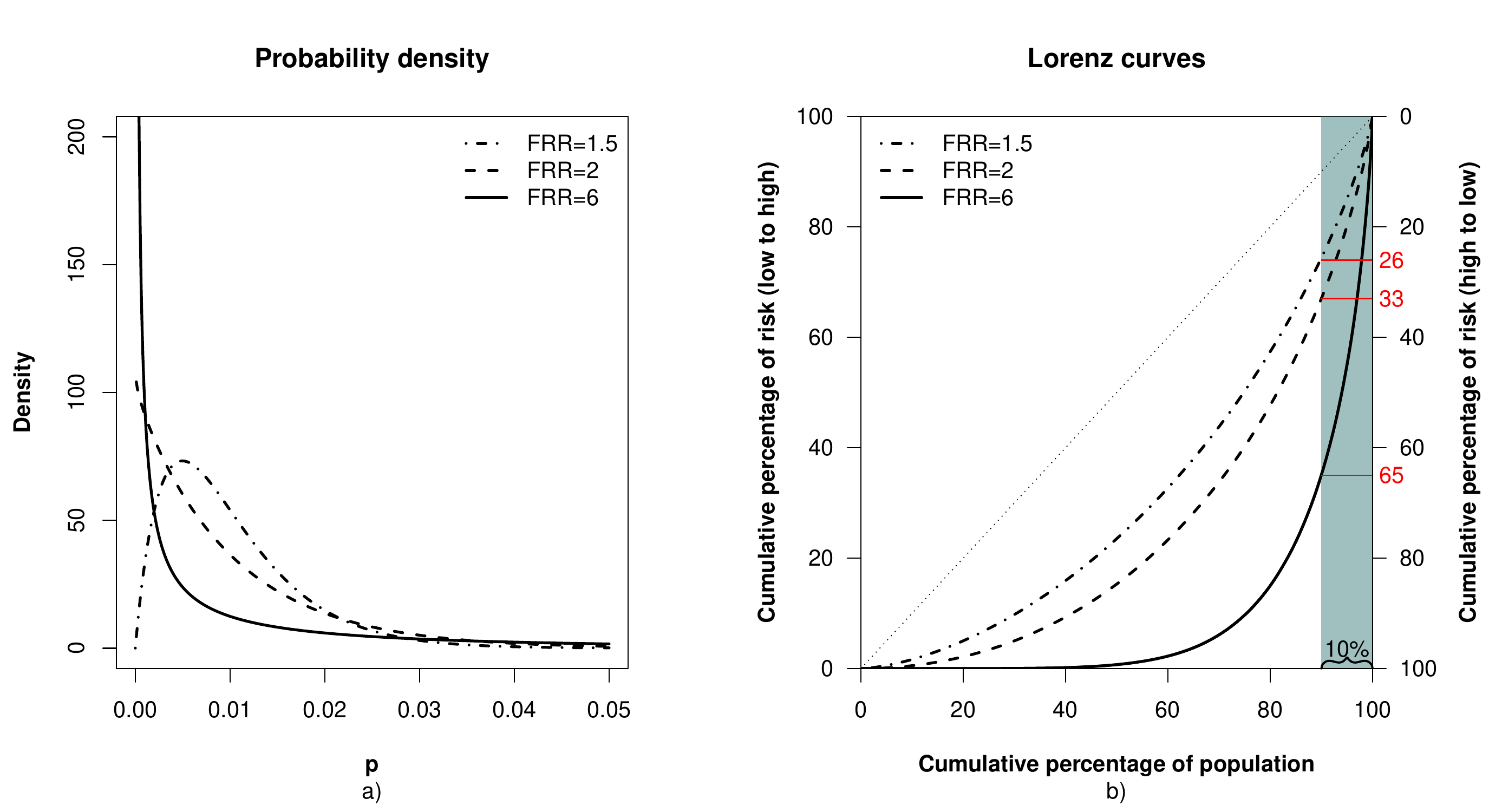}
 	\caption{The distribution (a) of a beta distributed variable $P$, and its corresponding Lorenz curve (b), for three selected familial relative risks ($FRR$s). The expected value of $P$ is set to 0.01. The 10\% at highest risk (shaded area) account for 26\%, 33\%, and 65\% of the diagnoses in the three scenarios, respectively.}
 	 \label{lorenz3eks}
\end{figure}
\end{landscape}



\section*{}
\ctable[
cap = solveBoth,
caption = Expressions \ref{FRR1} and \ref{FRR2} solved for the individual relative risk (IRR) and the high-risk proportion $q${,} according to the familial relative risks ($FRR$s) given one ($FRR_1$) and two ($FRR_2$) affected relatives. Estimates for $FRR_1$ and $FRR_2$ are taken from the indicated reference (the level of accuracy given in the reference is used).,
label = solveBoth
]{lcccccc}{
\tnote[]{Abbreviations: FRR, familial relative risk; IRR, individual relative risk; FDR, first degree relative.}
\tnote[a]{$\geq 2$ siblings/FDRs is used in the reference.}
}{ \FL
Cancer 												& Familial relationship studied	& $FRR_1$	& $FRR_2$		&$\Longrightarrow$		& $IRR$	& $q$ \\ \hline
Testicular cancer \cite{valberg2014hierarchical}			& Brothers					& 5.88		& 21.71			&$\Longrightarrow$		& 30.6	& 0.010\\  
Prostate cancer \cite{brandt2010age} 					& Brothers					& 2.96		& 7.71			&$\Longrightarrow$		& 12.2	& 0.027\\  	
Colorectal cancer \cite{johns2001systematic}				& Siblings					& 2.25		& 4.25\tmark		&$\Longrightarrow$		& 7.4	& 0.067\\  	
Melanoma \cite{fallah2014familial}						& FDRs						& 1.9		& 4.7\tmark		&$\Longrightarrow$		& 8.2	& 0.025\\  	
Breast cancer \cite{collaborative2001familial}			& Women FDRs				& 1.80		& 2.93			&$\Longrightarrow$		& 5.2	& 0.10\\  
Hodgkin's lymphoma \cite{kharazmi2015risk}			& FDRs						& 6			& 13\tmark		&$\Longrightarrow$		& 22.6	& 0.030\\ 
Non-medullary thyroid cancer \cite{fallah2013risk}		& FDRs						& 3.1		& 23.2\tmark		&$\Longrightarrow$		& 34.2	& 0.0022\\  \FL
}

%

\end{backmatter}
\end{document}